\documentclass[12pt]{article}

\usepackage{epsf}
\usepackage{amsmath}
\usepackage{amssymb}
\usepackage{graphicx}
\usepackage{dcolumn}
\usepackage{bm}

\begin{document}

\begin{center}
\Large\bf\boldmath
\vspace*{1.5cm}New constraints on supersymmetric models\\
from $b\to s\gamma$
\unboldmath
\end{center}

\vspace{0.6cm}
\begin{center}
F. Mahmoudi\footnote{Electronic address: \tt nazila.mahmoudi@tsl.uu.se}\\[0.4cm]
{\sl High Energy Physics, Department of Nuclear and Particle Physics, Uppsala University, \\
Box 535, 751 21 Uppsala, Sweden}
\end{center}
\vspace{0.6cm}
\begin{abstract}
\noindent We provide an analysis of the parameter space of several supersymmetry breaking scenarios such as the minimal supergravity (mSUGRA) model and the non universal Higgs mass (NUHM) framework, as well as the Anomaly Mediated Supersymmetry Breaking (AMSB) and the Gauge Mediated Supersymmetry Breaking (GMSB) models, in the light of a novel observable in $b\to s\gamma$ transitions, {\it i.e.} the isospin symmetry breaking in the exclusive $B\to K^*\gamma$ decays. We find that in many cases, this observable provides severe restrictions on the allowed parameter space regions for the mentioned models. Moreover, we provide a few examples of investigations of the physical masses of supersymmetric particles and search for the excluded values. The constraints from the branching ratio associated to $b\to s\gamma$ are also presented here for all the examined parameter space regions. A comparison with $B_s \to \mu^+ \mu^-$ branching ratio has also been performed.
\end{abstract}
\vspace{0.3cm}
\section{Introduction}
During the last few years, constraints from the branching ratio of $b\to s\gamma$ have been extensively used as a guide for supersymmetry phenomenology and in particular, to constrain the MSSM \cite{isidori-ellis}. Indeed, since these decays can only occur at loop level in the Standard Model, they bring very restrictive constraints on the new physics parameters.\\
In this study, we focus on a novel observable in $b\to s\gamma$ transitions, namely the isospin asymmetry, and show that this new observable can provide additional information to the inclusive branching ratio and, in some regions, even more restrictive limits on the SUSY parameters.\\
\\
The isospin asymmetry in the exclusive process $B\to K^*\gamma$ is defined as
\begin{equation}
\Delta_{0-}=\frac{\Gamma (\bar B^0\to\bar K^{*0}\gamma ) -\Gamma (B^-\to K^{*-}\gamma )}{\Gamma (\bar B^0\to\bar K^{*0}\gamma )+\Gamma (B^-\to K^{*-}\gamma
)}\;\; , \label{isospinasym}
\end{equation}
and similarly $\Delta_{0+}$ is defined as the charge conjugate of this equation. \\
Using QCD factorization, one can show that the isospin asymmetry can be written as \cite{kagan}:
\begin{equation}
\Delta_{0-}=\mbox{Re} (b_d-b_u)\;\; ,\label{asymb}
\end{equation}
where the spectator-dependent coefficients $b_q$ reads:
\begin{equation}
b_q = \frac{12\pi^2 f_B\,Q_q}{\bar m_b\,T_1^{B\to K^*} a_7^c}\left(\frac{f_{K^*}^\perp}{\bar m_b}\,K_1+ \frac{f_{K^*} m_{K^*}}{6\lambda_B m_B}\,K_{2q} \right)\;\;. \label{bq}
\end{equation}
In this equation, $a_7^c$, $K_1$ and $K_{2q}$ depend on the Wilson coefficients. We adopt here the definitions and conventions of \cite{ahmady} for the different parameters appearing in Eq. (\ref{bq}). An analysis of the branching ratio and isospin symmetry breaking in the context of beyond QCD factorization has also been performed in \cite{ball}.\\
\\
The experimental data for exclusive decays from Babar \cite{babar} and Belle \cite{belle} point to isospin asymmetries of at most a few percent, consistent with zero:
\begin{eqnarray}
\Delta_{0-}&=& +0.050 \pm 0.045({\rm stat.})\pm 0.028({\rm syst.})\pm 0.024(R^{+/0})\;\;\; (\mbox{Babar})\; , \label{babar}\\
\Delta_{0+}&=& +0.012 \pm 0.044({\rm stat.})\pm 0.026({\rm syst.})\;\;\; (\mbox{Belle})\; . \label{belle}
\end{eqnarray}
Calculating the expected isospin asymmetry from Eqs. (\ref{asymb}) and (\ref{bq}), and confronting the results to the combined experimental limits of (\ref{babar}) and (\ref{belle}) allow us to establish limits on the supersymmetry parameters. In \cite{ahmady}, we have detailed the calculation of the isospin asymmetry in the MSSM with minimal flavor violation, and performed scans on the mSUGRA parameter space. This study extends the analysis of \cite{ahmady} to a broader range of supersymmetric hypotheses, and we investigate the constraints from the isospin asymmetry for different scenarios of supersymmetry breaking. 
As a comparison reference, we also calculate the inclusive branching ratio associated to $b \rightarrow s \gamma$.\\
\\
All the calculations in this paper, for both the inclusive branching ratio and the isospin symmetry breaking have been performed with the computer program SuperIso \cite{superiso}, which is a public C program which calculates the isospin asymmetry, using a SUSY Les Houches Accord file for the input parameters, that can be either automatically generated thanks to for example SOFTSUSY \cite{softsusy} or ISAJET \cite{baer}, or provided by the user.\\
\\
In the following sections, after deriving the bounds on the isospin asymmetry and estimating the errors, we give a summary of the results in the minimal supergravity (mSUGRA) parameter space, and we present the constraints from isospin asymmetry for other scenarios such as the non universal Higgs mass (NUHM), the Anomaly Mediated Supersymmetry Breaking (AMSB) and the Gauge Mediated Supersymmetry Breaking (GMSB) models.\\
\section{Bounds on isospin asymmetry}
%
\begin{table}[!t]
\begin{center}
\begin{tabular*}{160mm}{@{\extracolsep\fill}|c|c|c|c|c|}
\hline\hline
\multicolumn{5}{|c|}{CKM parameters and $B$ meson mass}\\
\hline
$V_{us}$ & $V_{cb}$~~~ & $\left|V_{ub}/V_{cb}\right|$~~ & $\mbox{Re}(V_{us}^* V_{ub}/V_{cs}^*V_{cb})$~~~ & $m_B$~~~ \\
\hline
~~~~0.22~~~~ & $0.041 \pm 0.05~~~$ & $0.085 \pm 0.025$ & $0.011\pm 0.005~~~$ & 5.28 GeV~~~\\
\hline
\end{tabular*}
\\
\vspace*{0.1cm}
\begin{tabular*}{160mm}{@{\extracolsep\fill}|c|c|c|c|c|}
\hline
\multicolumn{5}{|c|}{$B$ meson parameters}\\
\hline
$f_B$ & $\lambda_B$~~ & $ a_{\perp} $~~ & $ < \bar v^{-1} >_\perp $~ & $h_{K^*}(x)~$\\
\hline
$200 \pm 20$ MeV & $350 \pm 150$ MeV~ & $ 0.19 \pm 0.02 $~~ & $ 3.7 \pm 0.04$~~ & $ ^{\displaystyle(4.8 \pm 0.5)~~}_{\displaystyle ~~+ (1.5 \pm 0.2)i} $ \\
\hline
\end{tabular*}
\\
\vspace*{0.1cm}
\begin{tabular*}{160mm}{@{\extracolsep\fill}|c|c|c|c|}
\hline
\multicolumn{4}{|c|}{$K^*$ meson parameters}\\
\hline
~~~~$f_{K^*}^\perp$~~ & $m_{K^*}$~~~~~ &  $f_{K^*}$~~~~~~ & $T_1^{B\to K^*}$~~~~~ \\
\hline
~~~~$175 \pm 9$ MeV~~~ & 892 MeV~~~~~~ & $226 \pm 28$ MeV~~~~~~ & $0.30 \pm 0.05$~~~~~~\\
\hline
\end{tabular*}
\\
\vspace*{0.1cm}
\begin{tabular*}{160mm}{@{\extracolsep\fill}|c|c|c|c|}
\hline
\multicolumn{4}{|c|}{Convolution integral parameters}\\
\hline
$F_\perp$ & $G_\perp(x_{cb})$ & $H_\perp(x_{cb})$~~ & $X_\perp$~~\\
\hline
~~~~~$1.21\pm 0.06$~~~~~ & $^{\displaystyle(2.82\pm 0.20)~~~}_{\displaystyle~~~+(0.81\pm 0.23)i}$ & $^{\displaystyle(2.32\pm 0.16)~~~}_{\displaystyle~~~+(0.50\pm 0.18)i}$ & $^{\displaystyle(3.44\pm 0.47)\,X~~~}_{\displaystyle~~~-(3.91\pm 1.08)}$ \\
\hline
\end{tabular*}
\\
\vspace*{0.1cm}
\begin{tabular*}{160mm}{@{\extracolsep\fill}|c|c|c|c|c|}
\hline
\multicolumn{5}{|c|}{Quark and $W$-boson masses}\\
\hline
$m_b(m_b)$ & $m_c(m_b)$~ & $m_s$~~ & $m_t$~~ & $M_W$~\\
\hline
$4.2 \pm 0.1$ GeV & $1.2 \pm 0.2$ GeV & $0.10 \pm 0.03$ GeV & $172.5 \pm 2.7$ GeV & 80.4 GeV\\
\hline\hline
\end{tabular*}
\end{center}
\caption{Numerical values of the parameters involved in the calculation of the isospin asymmetry. \vspace*{1.5cm}}
\label{tab:param}
\end{table}%
\begin{center}
\begin{figure}[ht!]
\begin{center}\includegraphics[width=7.5cm]{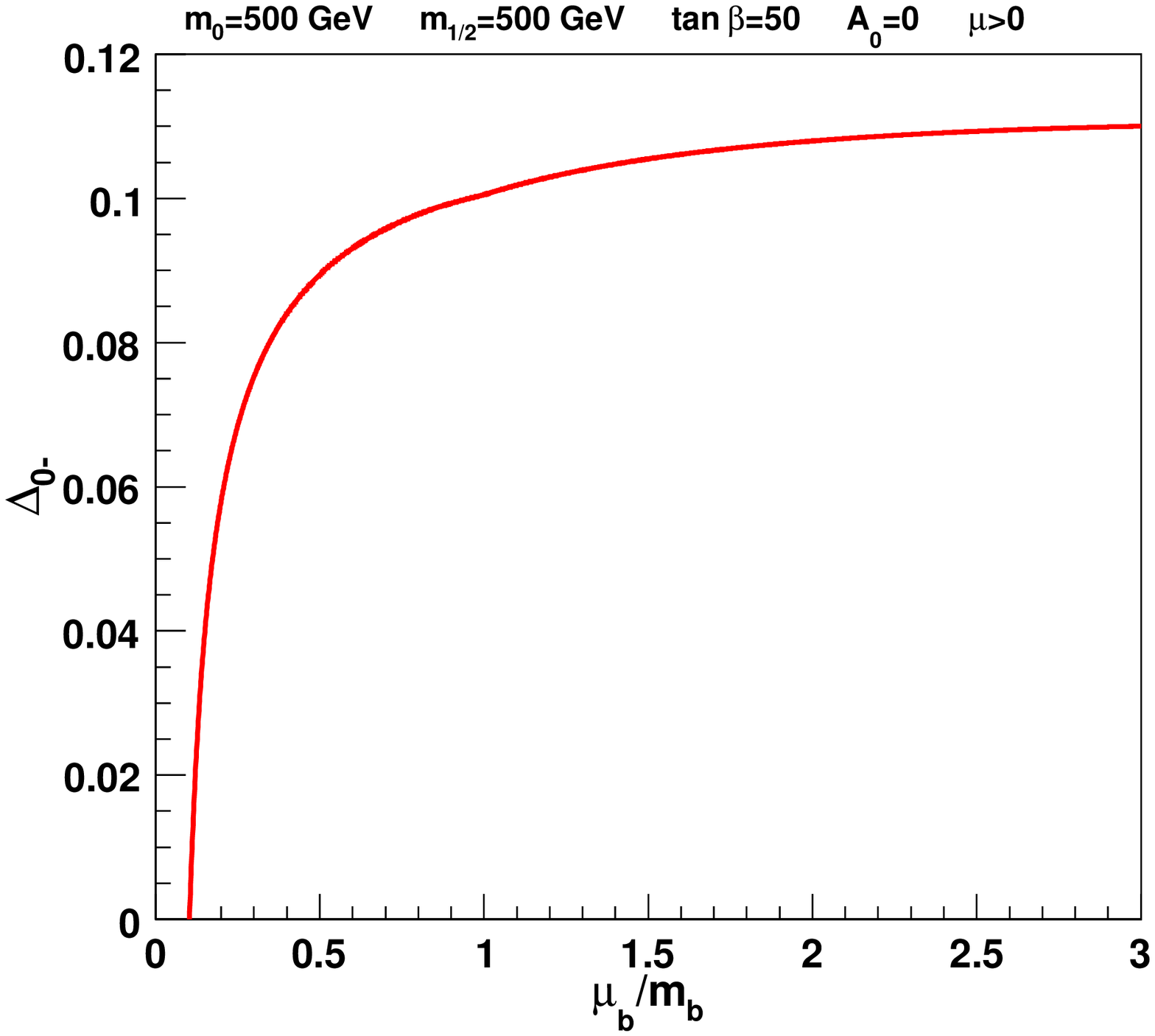}\includegraphics[width=7.5cm]{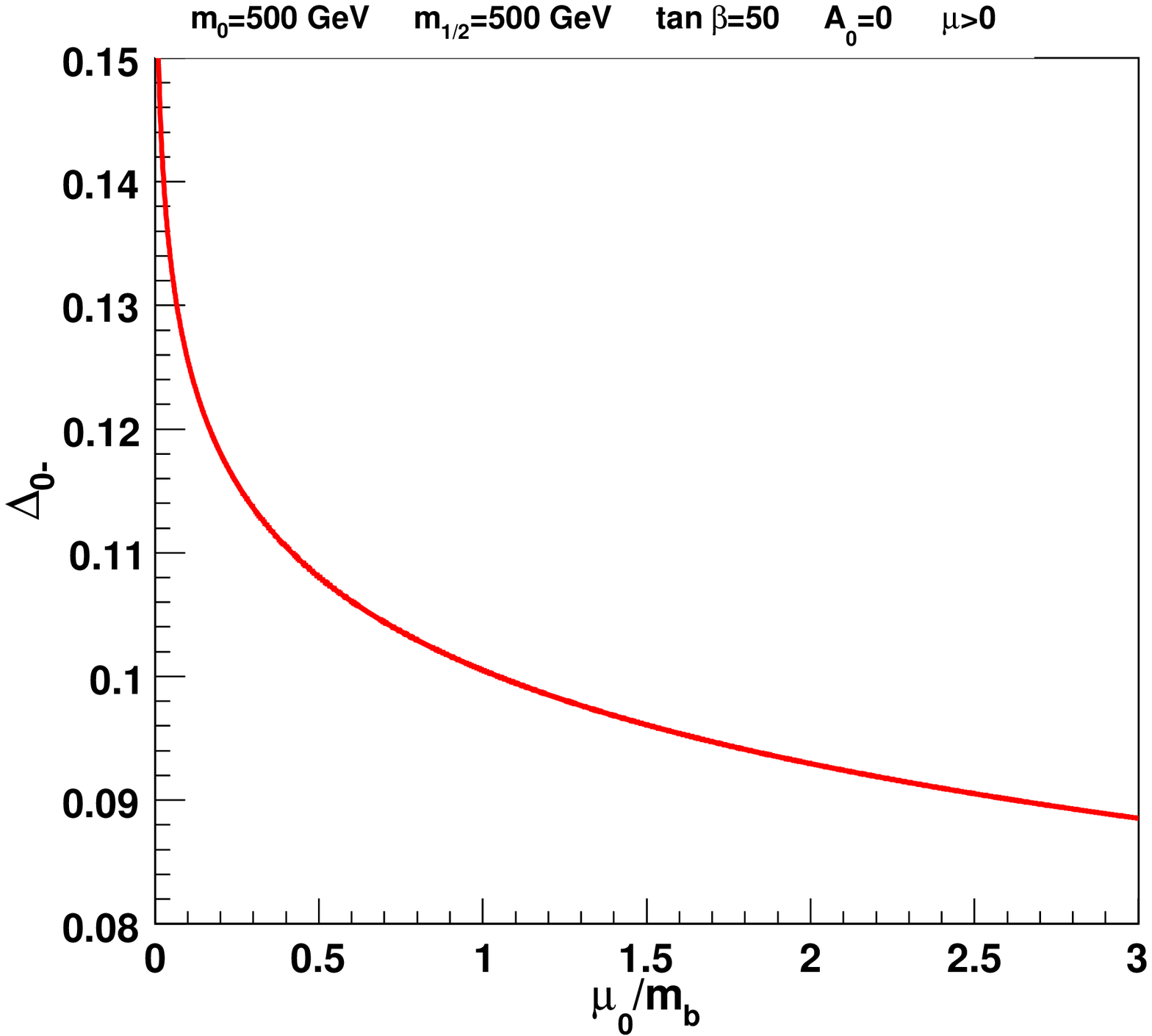}\\
\includegraphics[width=7.5cm]{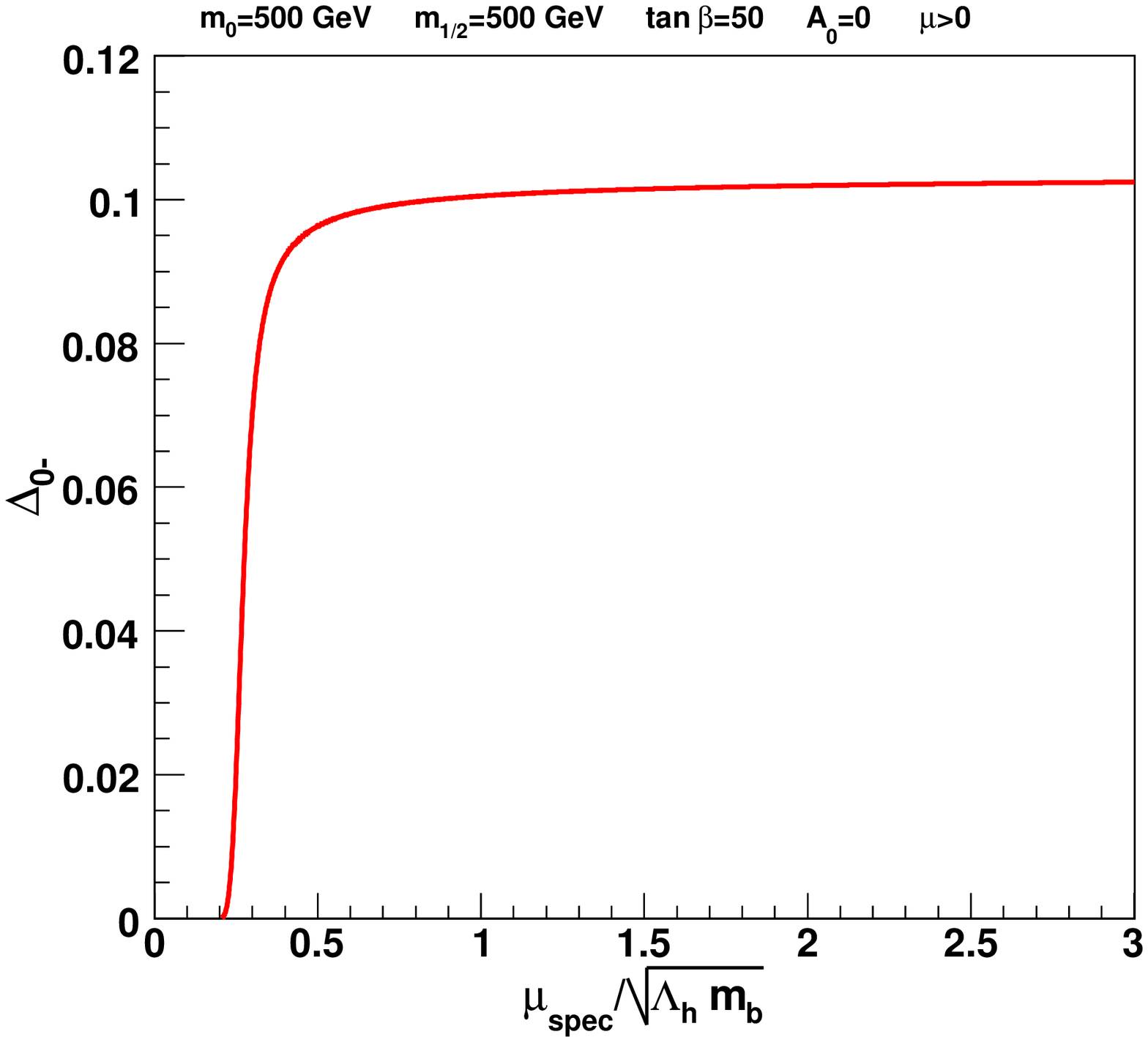}
\end{center}
\caption{Dependence of the isospin asymmetry on the scales $\mu_b$, $\mu_0$ and $\mu_{spec}$. We consider here the mSUGRA parameter space with $m_0=500$ GeV, $m_{1/2}=500$ GeV, $\tan \beta=50$, $A_0=0$ and $\mu>0$. \vspace*{1.0cm}}
\label{scales}
\end{figure}
\end{center}%
\begin{figure}[ht!]
\begin{center}
\includegraphics[width=10cm]{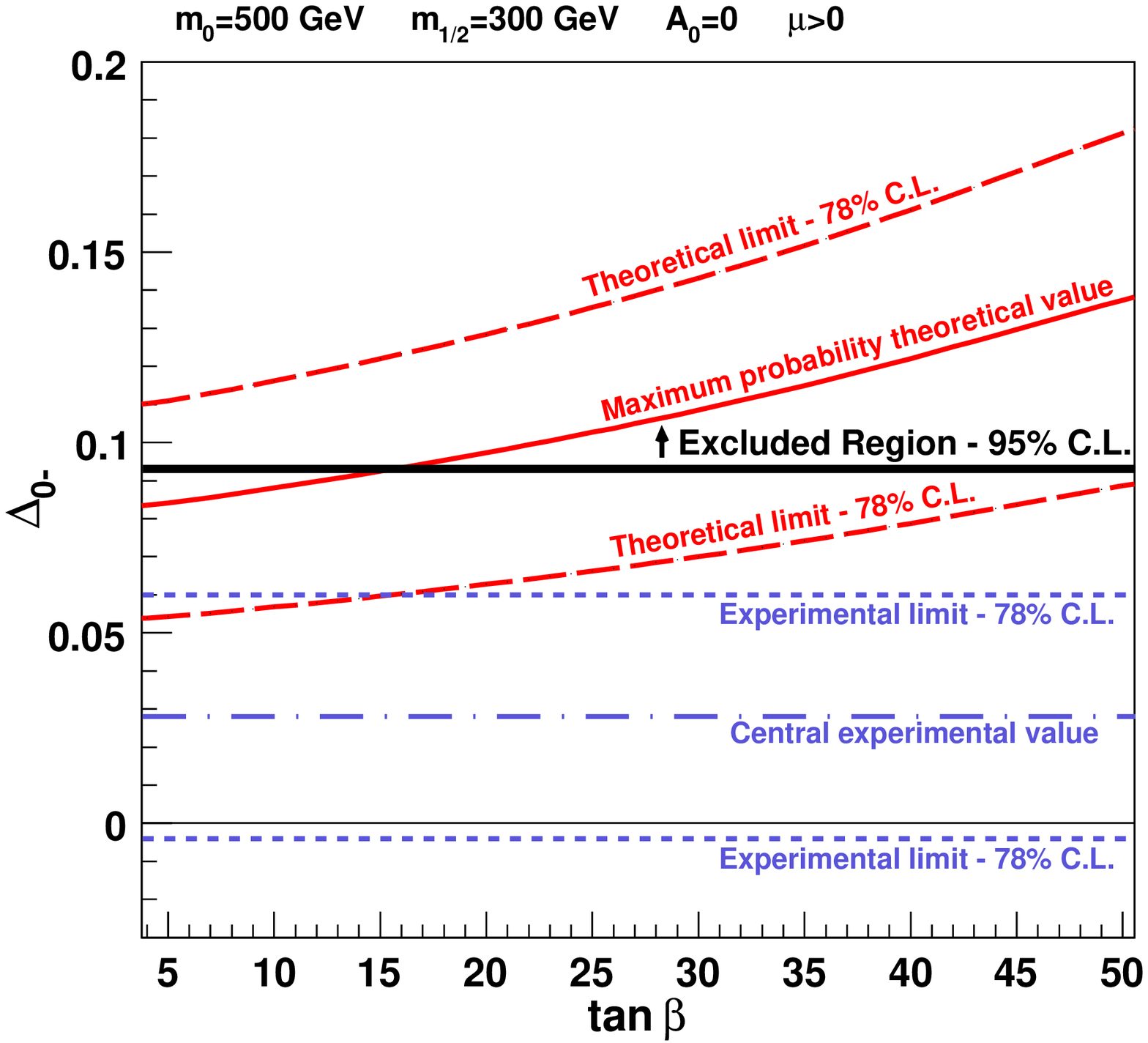}
\caption{Evaluation of the experimental and theoretical errors, in the mSUGRA parameter space with $m_0=500$ GeV, $m_{1/2}=300$ GeV, $A_0=0$ and $\mu>0$. The horizontal black line corresponds to the criterion (2.1), which is a 95\% confidence level limit, to be compared with the red plain line corresponding to the calculated isospin asymmetry. \vspace*{1.0cm}}
\label{fig:limit}
\end{center}%
\end{figure}%
In this section, we perform a general analysis of the errors, in order to derive the effective bounds on the isospin asymmetry.\\
\\
The calculation of $\Delta_{0-}$ requires the knowledge of many parameters (please refer to \cite{kagan,ahmady} for a complete description of the calculation), whose values and associated errors are given in Table~\ref{tab:param}\footnote[1]{Most of the values in this table are taken from \cite{bosch} and \cite{ali}, with some updates.}. The parameter $\displaystyle X=\ln(m_B/\Lambda_h)\,(1+\varrho\,e^{i\varphi})$ in this table parametrizes the logarithmically divergent integral $\int_0^1 dx/(1-x)$. Allowing as usual $\varrho \le 1$  and an arbitrary phase $\varphi$, we perform an analysis of the errors due to the variation of the input parameters of Table~\ref{tab:param}. We find, at 95\% C.L., that the total relative theoretical error is about 35\%. The highest relative uncertainties arise from $\lambda_B$ (10\%), $T_1^{B \rightarrow K^*}$ (7\%), $f_B$ (3\%), and $X$ (3\%).\\
\\
However, the isospin asymmetry calculation also involves the choice of three different scales, $\mu_b = O(m_b)$, $\mu_0 = O(m_b)$ and $\mu_{spec} = O(\sqrt{\Lambda_h m_b})$, where $\Lambda_h$ is a hadronic scale that we take to be approximately 0.5 GeV. The dependence of the theoretical predictions on the choice of the scales can be considered as an estimate of higher-order corrections. This dependence is depicted in Fig.~\ref{scales}.
We can first remark that $\Delta_{0-}$ is quite stable with respect to the variation of $\mu_{spec}$. We can also observe a higher scale dependence of $\Delta_{0-}$ for small values of the $\mu$'s. Following the usual practice, we evaluate the truncation errors while varying the scales $\mu_b$, $\mu_0$ and $\mu_{spec}$ independently, in the ranges $\mu_b \in [m_b/2,2 m_b]$, $\mu_0 \in [m_b/2,2 m_b]$ and $\mu_{spec} \in [\sqrt{\Lambda_h m_b}/2,2 \sqrt{\Lambda_h m_b}]$, with their central values taken to be respectively $m_b$, $m_b$ and $\sqrt{\Lambda_h m_b}$. We then calculate the truncation error of the sum by adding the individual errors in quadrature. We determine the relative uncertainty corresponding to the choice of the scales: -15\% / + 10\%. This error can be considered as an evaluation of the influence of higher order contributions.
Combining all the sources of errors, we find that the total relative theoretical uncertainty at 95\% C.L. is -50\% / +45\%.\\
\\
Combining the experimental and the theoretical errors, we derive
the criterion (at 95\% C.L.):
\begin{equation}
-0.018 < \Delta_{0-} < 0.093 \;\;.
\label{limit} 
\end{equation}%
This criterion is illustrated in Fig.~\ref{fig:limit}. The upper limit will be used in the following sections to impose constraints on supersymmetric parameter spaces. Although the uncertainties are very large, we can safely claim that the theoretical predictions for the isospin asymmetry should not exceed 9.3\% (at 95\% C.L.), and this allows us to rule out models (or parameters) which produce too large isospin asymmetries.\\
\newpage
\section{Isospin asymmetry and the MSSM}
In this section, a comparison between the theoretical evaluations and the experimental bounds of the last section has been performed. We investigate the constraints from the isospin asymmetry for several scenarios of supersymmetry breaking.\\
\subsection{mSUGRA}
A more detailed investigation of the minimal supergravity model (mSUGRA) parameter space has been presented in \cite{ahmady,mahmoudi}. Here we emphasize the major results and provide some comparisons with other $B$ Physics observables.\\
\\
\begin{figure}[!hb]
\hspace*{-0.5cm}
\includegraphics[width=8.7cm,height=8cm]{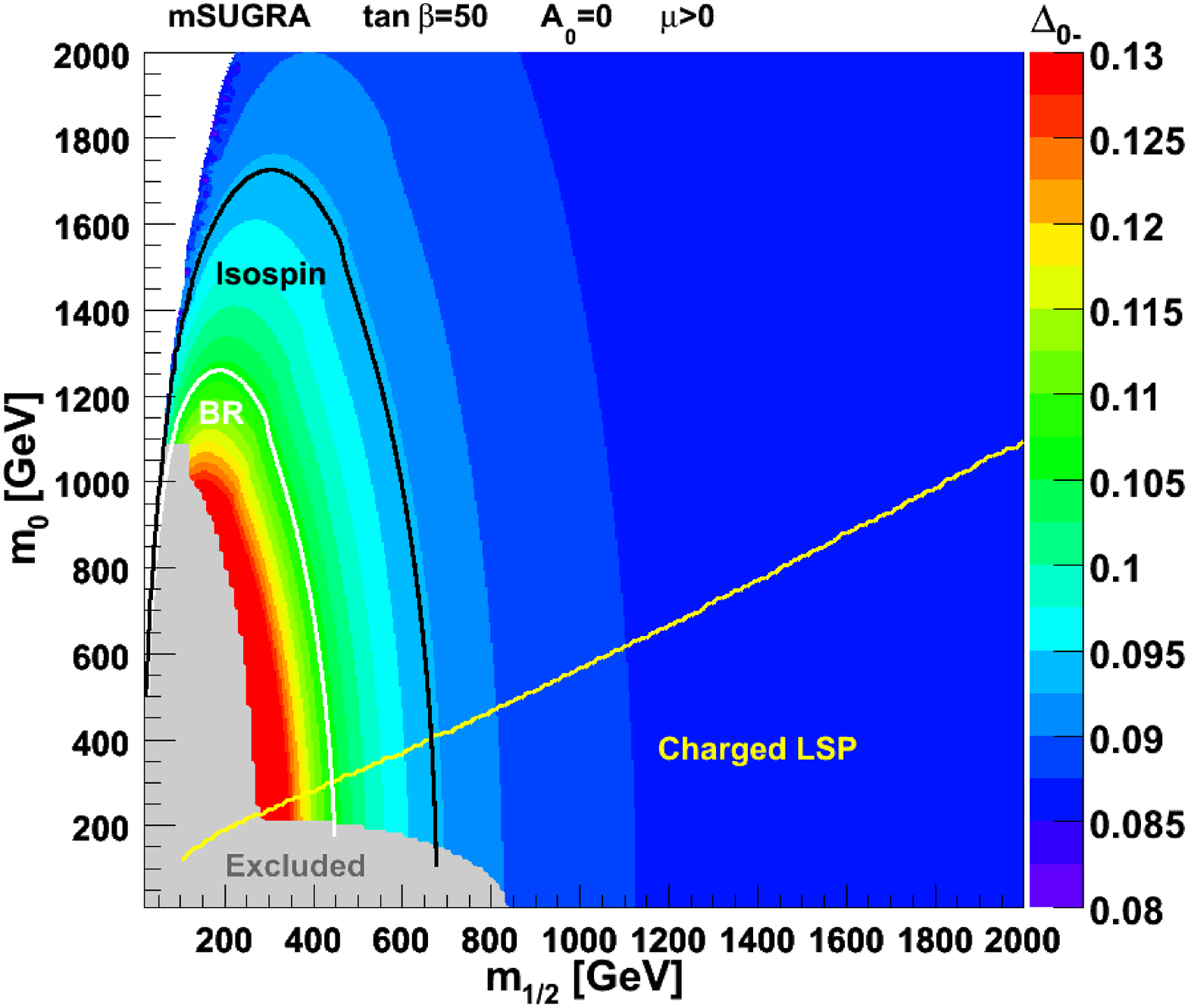}~~\includegraphics[width=8.7cm,height=8cm]{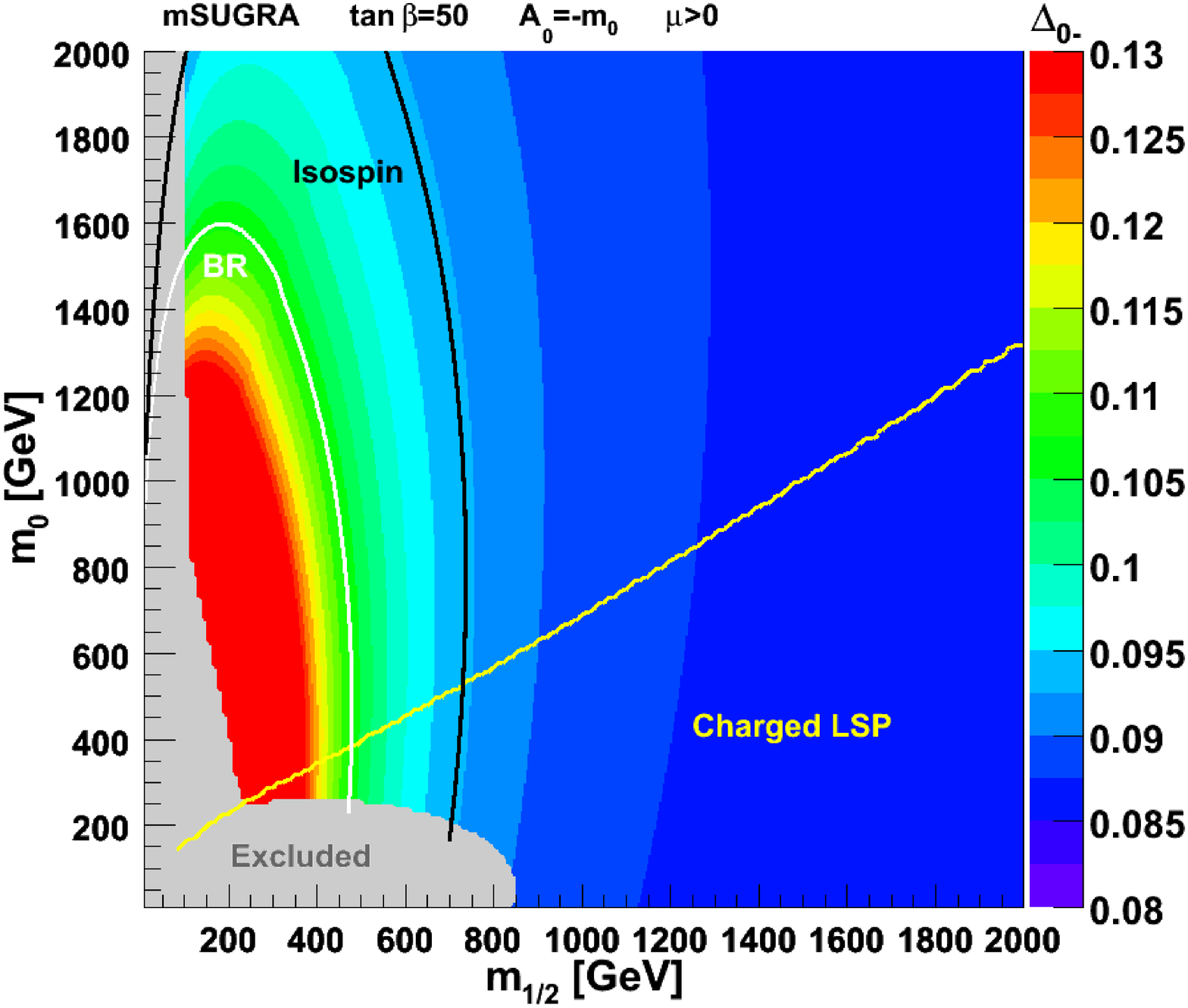}\\
\caption{Constraints on the mSUGRA parameter plane $(m_{1/2},m_0)$ for $A_0=0$ (left) and $A_0=-m_0$ (right). The conventions for the colors and the meaning of the different regions are described in the text. \vspace*{1.0cm}}
\label{msugra}
\end{figure}\\
The SUSY mass spectra, as well as the couplings and the mixing matrices were generated using SOFTSUSY 2.0.14 \cite{softsusy}. \\
\\
\begin{figure}[!ht]
\begin{center}
\includegraphics[width=9.0cm,height=8cm]{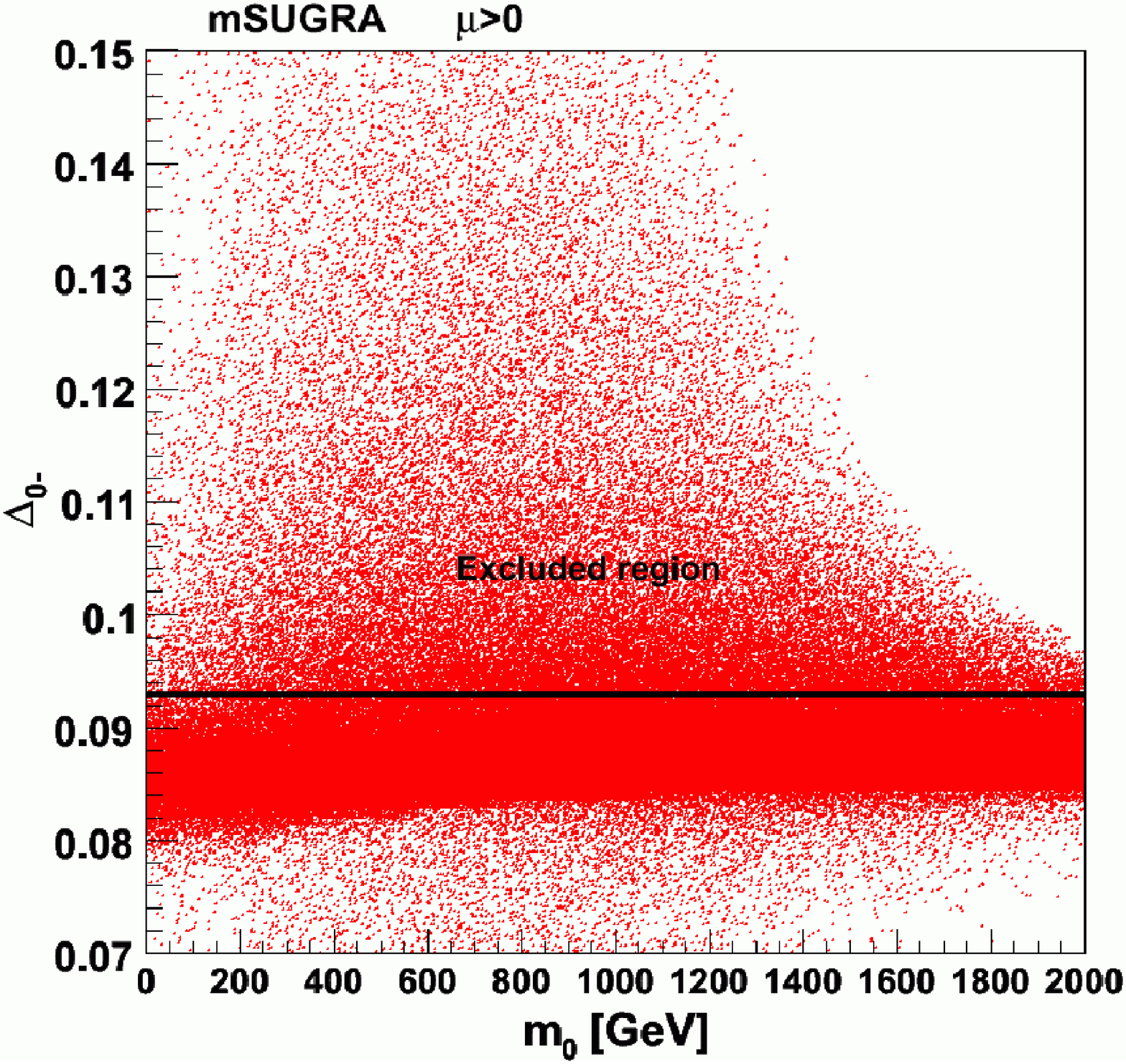}\includegraphics[width=9.0cm,height=8cm]{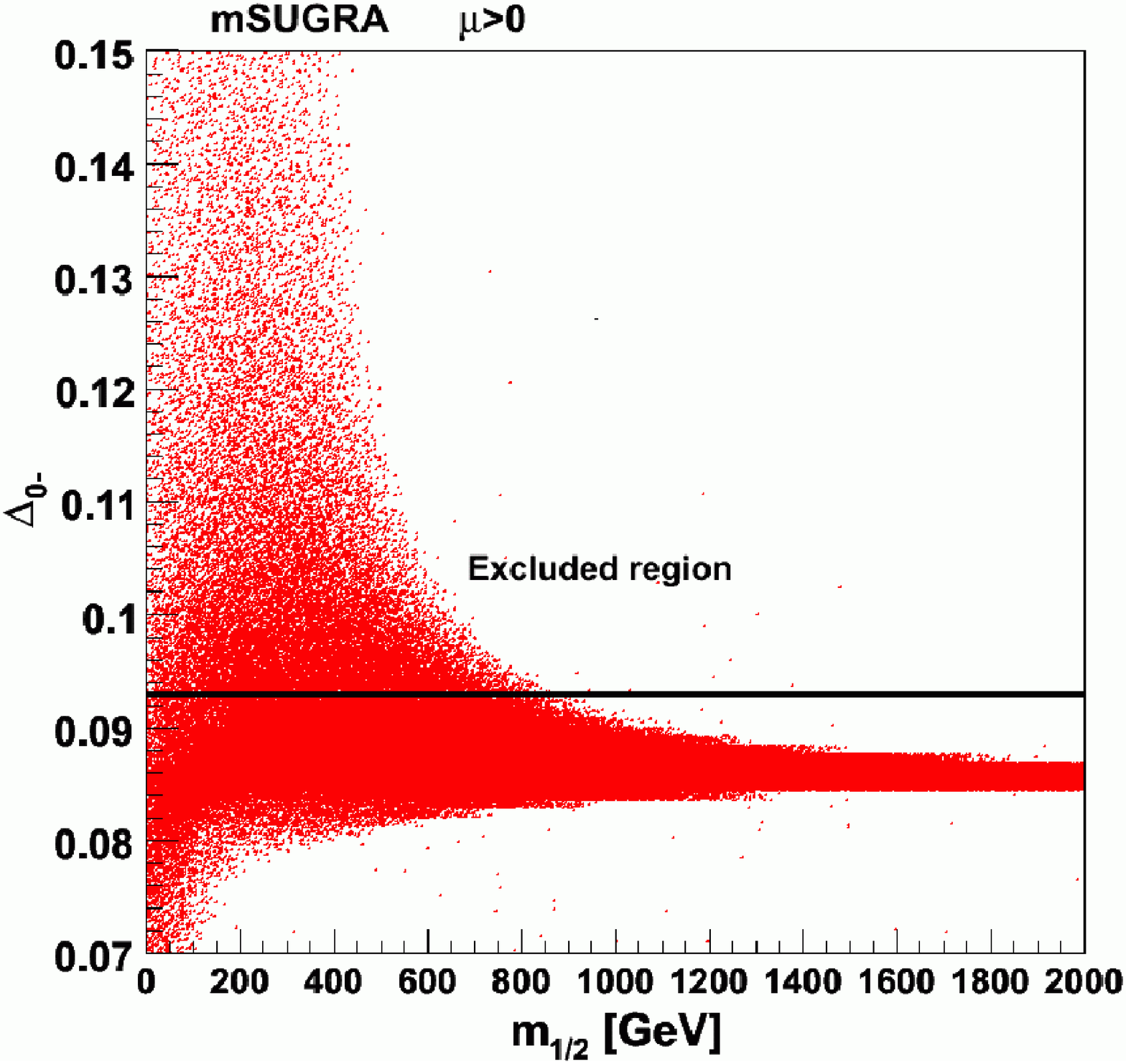}\\
\vspace*{0.5cm}
\includegraphics[width=9.0cm,height=8cm]{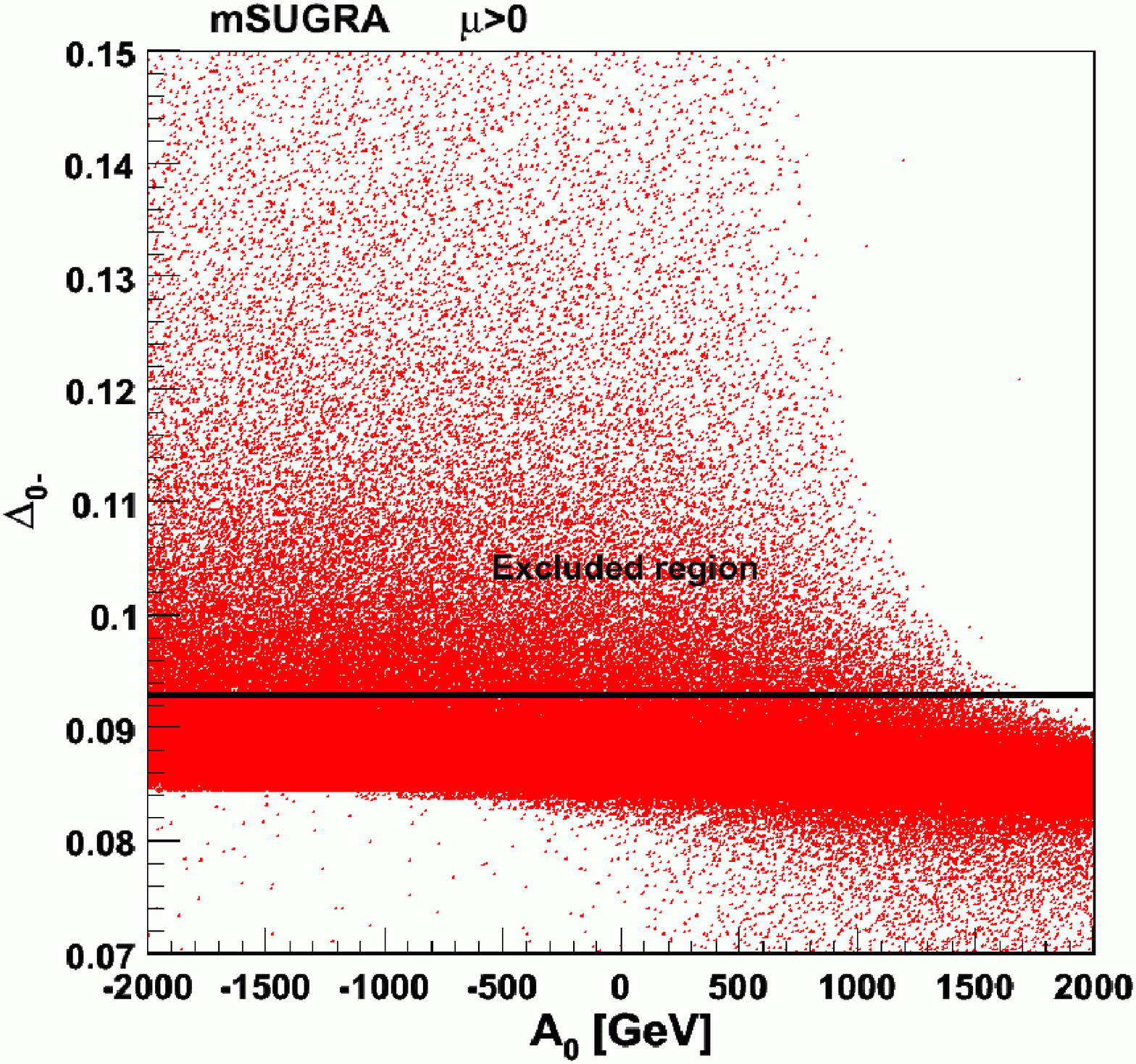}\includegraphics[width=9.0cm,height=8cm]{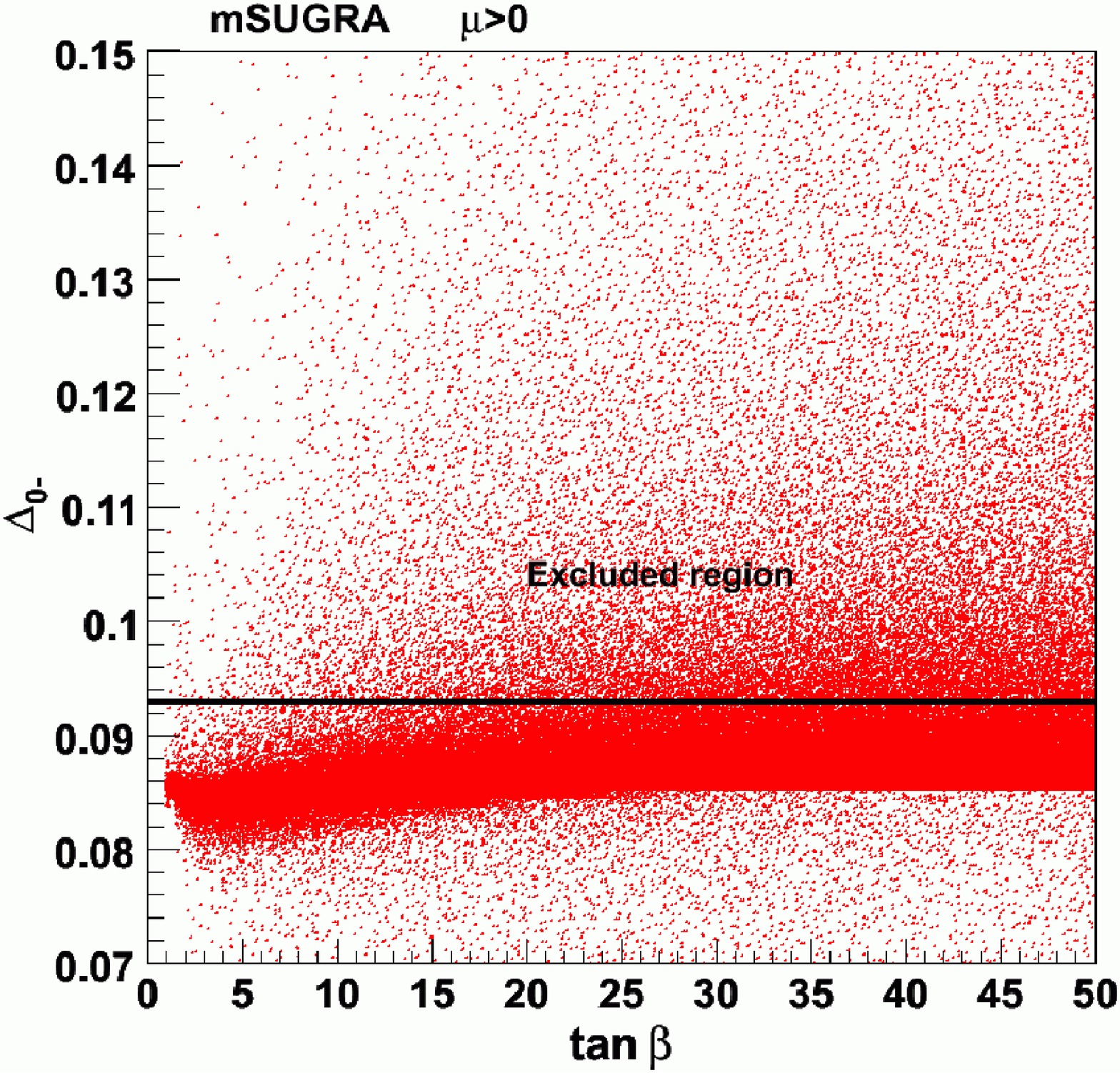}
\caption{Dependence of isospin symmetry breaking on the different input parameters of the mSUGRA model, when scanning over one million randomly chosen parameter space points, for $\mu > 0$.}
\label{tout}
\end{center}
\end{figure}%
\begin{figure}[!t]
\begin{center}
\includegraphics[width=8.7cm,height=8cm]{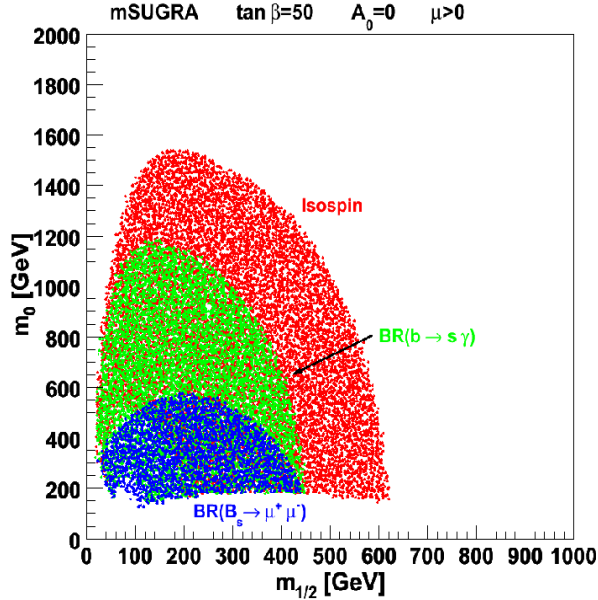}\\
\caption{Excluded regions in the mSUGRA parameter plane $(m_{1/2},m_0)$ for $\tan\beta=50$ and $A_0=0$. The red region is excluded by the isospin asymmetry of $B \to K^* \gamma$, while the green region is excluded by the inclusive branching ratio of $B \to X_s \gamma$. The blue region is excluded by the branching ratio of $B_s \to \mu^+ \mu^-$. Note that the red region contains the green one which contains the blue region.}
\label{mu}
\end{center}
\end{figure}%
We scan the mSUGRA parameter space $\lbrace m_0, m_{1/2}, A_0, \tan\beta, \mathrm{sign}(\mu) \rbrace$, and for every point we calculate the isospin asymmetry and confront it to the limits of Eq. (\ref{limit}). We also calculate the inclusive branching ratio as a comparison reference. Considering the latest experimental limits from the Heavy Flavor Averaging Group (HFAG) \cite{hfag}, including theoretical errors from \cite{misiak,becher}, as well as an intrinsic MSSM correction \cite{BR}, we derive the following limits at 95\% C.L.:
\begin{equation}
2.07 \times 10^{-4} < \mathcal{B}(b \to s \gamma) < 4.84 \times 10^{-4} \;\;.
\end{equation}\\
In Fig.~\ref{msugra}, an investigation of the $(m_{1/2},m_0)$ plane for $A_0=0$ and $A_0=-m_0$ is presented. In this figure, inside the black contour marked ``Isospin'' is excluded by the isospin breaking constraints, whereas inside the contour marked ``BR'' corresponds to the region excluded by the inclusive branching ratio constraints. The ``Excluded'' light grey area in the figure corresponds to the case where at least one of the sparticle masses does not satisfy the collider constraints or where the neutral Higgs boson becomes too light \cite{PDG2006}. Finally, the ``Charged LSP'' region is cosmologically disfavored if R-parity is conserved. The various colors represent the changing magnitude of the isospin asymmetry.\\
One can notice the severe constraints from the isospin symmetry breaking for large $\tan\beta$ values and for $\mu > 0$ as compared to the total branching ratio. One can also note that the isospin asymmetry is enhanced by a negative value of $A_0$, although the global shapes of the limiting regions remain similar.\\
\\
In order to have a better idea of the dependence of isospin asymmetry on different mSUGRA parameters, we present in Fig.~\ref{tout} the results of the scan over one million randomly chosen parameter space points while varying the mSUGRA input parameters in the ranges $m_0 \in [0,2000]$, $m_{1/2} \in [0,2000]$, $A_0 \in [-2000,2000]$ and $\tan \beta \in [0,50]$, for $\mu > 0$. The horizontal black line in these plots corresponds to the limit of Eq. (\ref{limit}). One can notice here a larger number of excluded points for higher values of $\tan\beta$, small values of $m_{1/2}$ and negative values of $A_0$. Approximately 10\% of the analyzed points are in the excluded region.\\
\\
To evaluate how restrictive the isospin symmetry breaking is compared to the other $B$ Physics observables, we show in Fig.~\ref{mu} an example of the regions excluded by the branching ratio of $B_s \to \mu^+ \mu^-$ (in blue), by the inclusive branching ratio of $B \to X_s \gamma$ (in green) and by the isospin asymmetry of $B \to K^* \gamma$ (in red), for $\tan\beta=50$ and $A_0=0$. \\
We can remark that isospin asymmetry is more constraining than both the branching ratio observables.\\
For this plot, we used the following constraint for the branching ratio of $B_s \to \mu^+ \mu^-$ \cite{mu}:
\begin{equation}
\mathcal{B}(B_s \to \mu^+ \mu^-) < 0.97 \times 10^{-7} \;\;,
\end{equation}%
and the masses and couplings were generated using ISAJET 7.75 \cite{baer}.\\
\\
In this section, we showed that in the studied mSUGRA regions, the isospin asymmetry greatly enlarges the exclusion contours compared to the previously used $B$ physics observables.\\
\\
\begin{figure}[!t]
\begin{center}
\includegraphics[width=8.7cm,height=8cm]{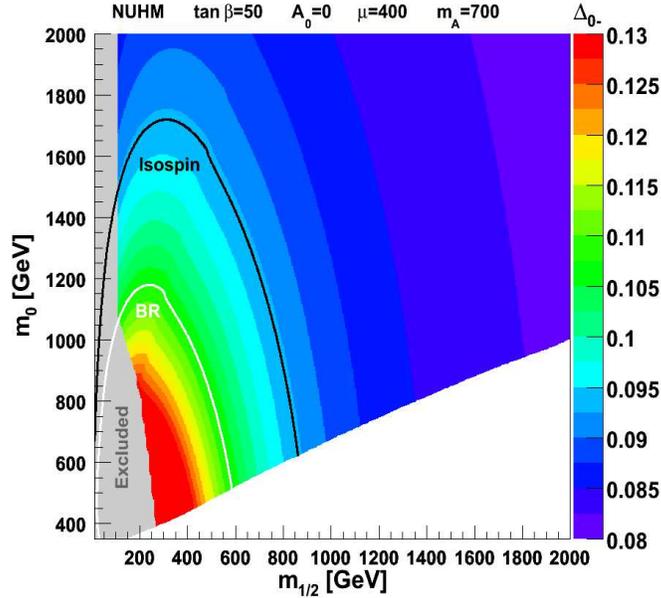}\\
\caption{Constraints on the NUHM parameter plane $(m_{1/2},m_0)$. The conventions for the different colored regions are the same as in Fig. 3. In the white region tachyonic particles are encountered. \vspace*{0.5cm}} \label{nuhm-m0m1/2}
\end{center}
\end{figure}%
\subsection{NUHM}
We explore in this section the non universal Higgs mass (NUHM) framework parameter space \cite{nuhm}, in which the universality assumptions of the soft SUSY breaking contributions to the Higgs masses are relaxed as compared to the mSUGRA scenario. Within this framework, two additional free parameters, $M_A$ and $\mu$, add to the five universal parameters of the mSUGRA scenario.\\
\\
Fig.~\ref{nuhm-m0m1/2} shows an example of the results in the $(m_{1/2},m_0)$ plane for $\tan\beta=50$, $A_0=0$, $M_A=700$ GeV and $\mu=400$ GeV. The masses and couplings were generated using SOFTSUSY 2.0.14 \cite{softsusy}. The results are similar to those for the mSUGRA parameter space, as was expected. The white area at the lower part of this figure has not been generated since it corresponds to tachyonic particles.\\
\\
\begin{figure}[!t]
\hspace*{-0.5cm}
\includegraphics[width=8.7cm,height=8cm]{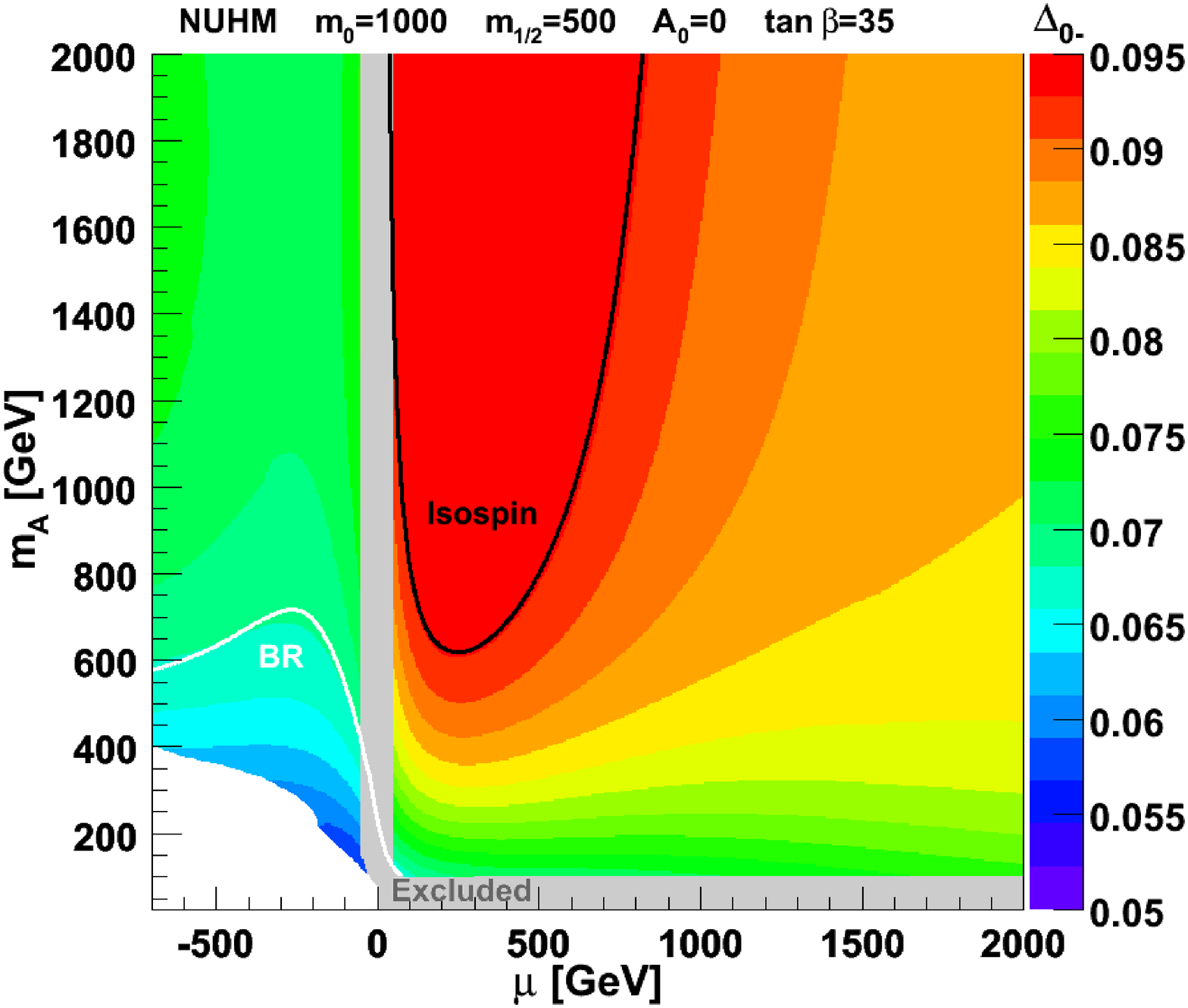}~~\includegraphics[width=8.7cm,height=8cm]{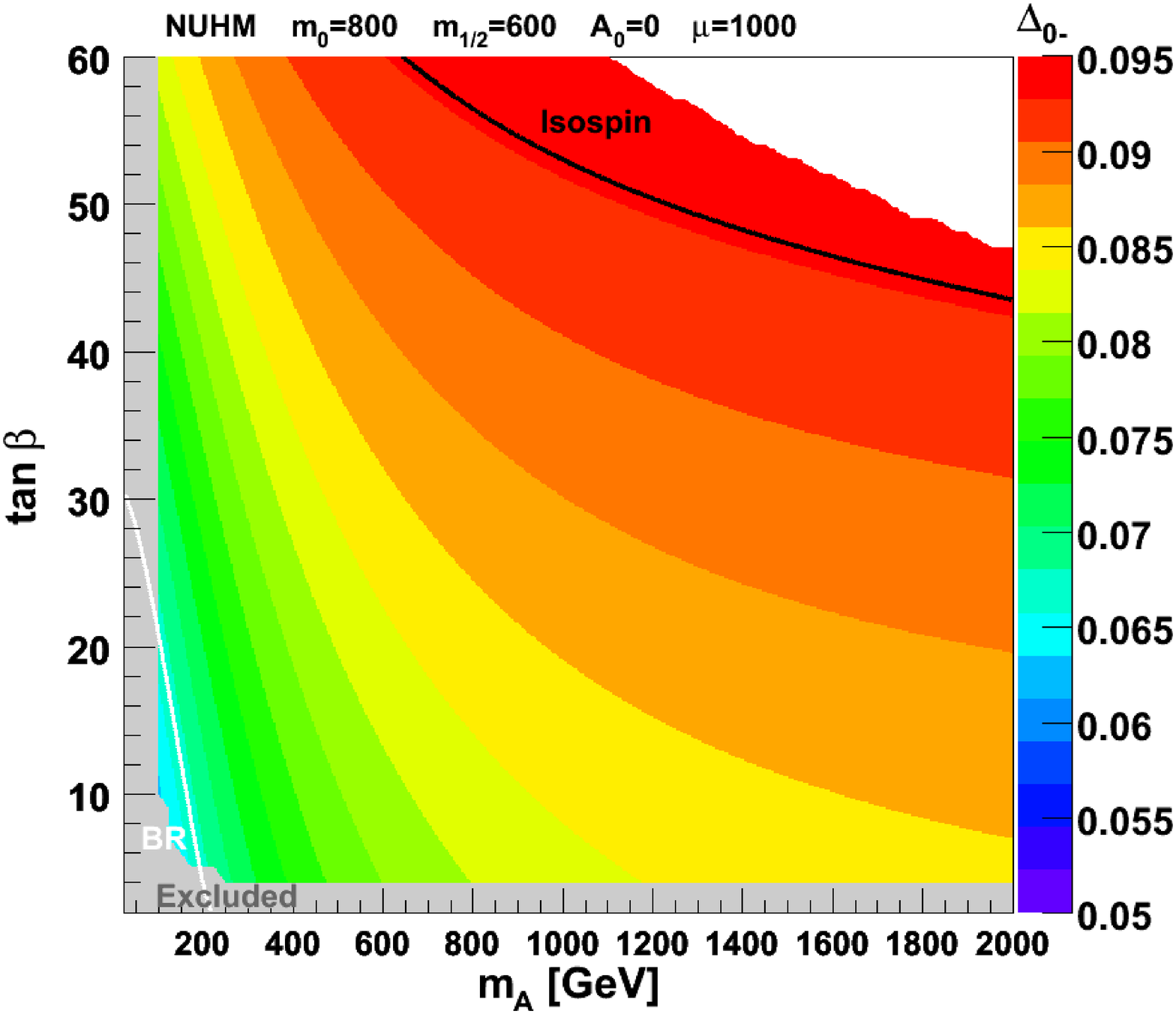}\\
\caption{Constraints on the NUHM parameter planes, $(\mu,m_A)$ to the left and $(m_A,\tan\beta)$ to the right. The white BR contour delimits the region excluded by the branching ratio, and the black contour corresponds to the isospin symmetry breaking constraint. The conventions for the different regions are the same as in the precedent figures, but the color scale here is different. \vspace*{0.6cm}}
\label{nuhm}
\end{figure}%
In Fig.~\ref{nuhm}, the $(\mu,m_A)$ and $(m_A,\tan\beta)$ planes are investigated. For these two samples, the regions excluded by the branching ratio and by the isospin asymmetry are not correlated anymore. One can thus appreciate the additional information provided by this new observable. Furthermore, these results can be compared to other existing constraints, for example those from WMAP. For instance, comparing the $(\mu,m_A)$ plane (Fig.~\ref{nuhm}) with a similar plot presented in \cite{ellis}, one can notice that the WMAP favored region was between two strips at roughly constant positive and negative values of $\mu$, extending approximately to $\mu =350$ GeV. It is remarkable that the isospin asymmetry constraint reduces a substantial part of this region. This example illustrates the usefulness of exploring isospin asymmetry and the complementary information that can be obtained.\\
\\
\subsection{AMSB}
\begin{figure}[!t]
\hspace*{-0.5cm}
\includegraphics[width=8.7cm,height=8cm]{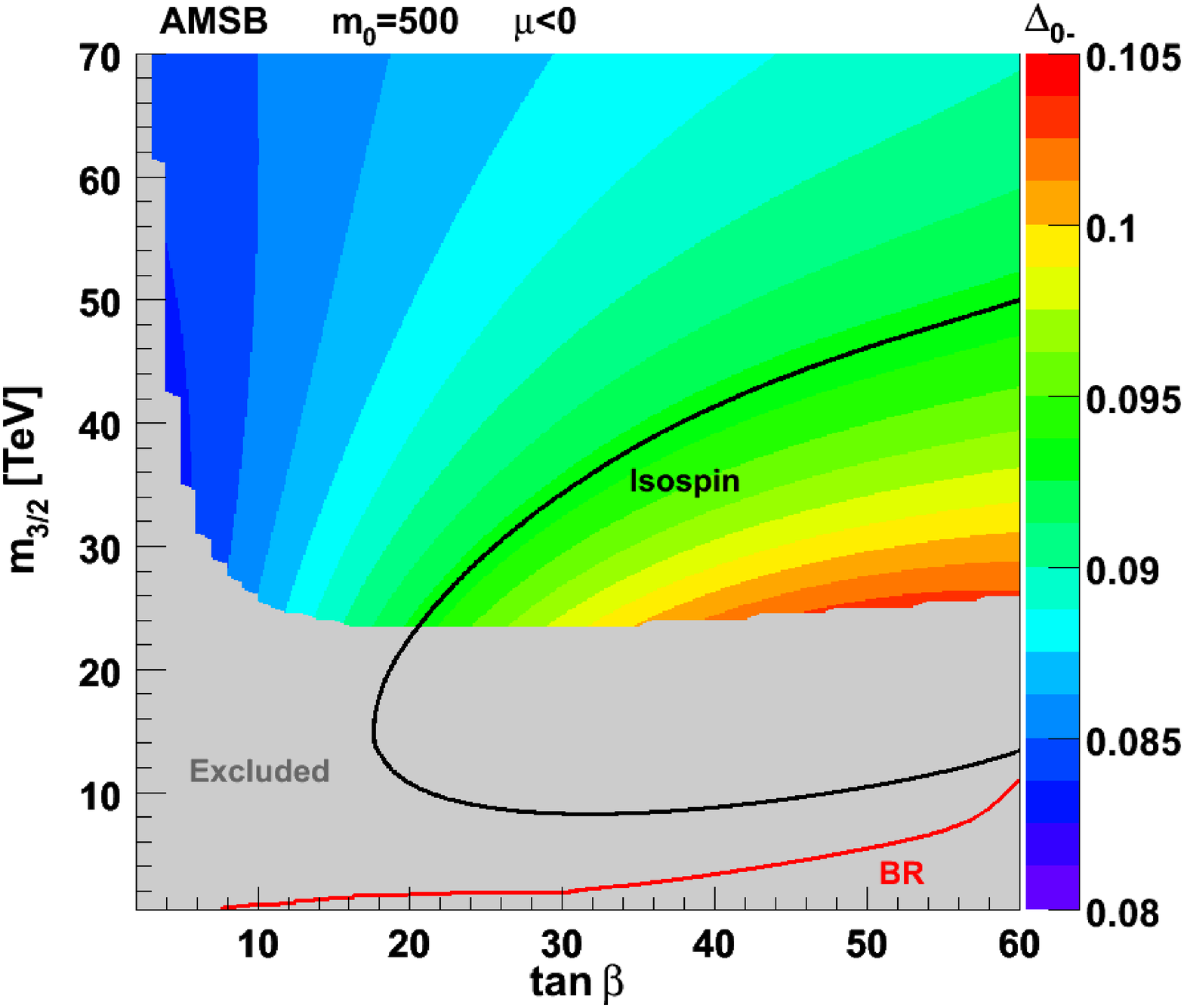}~~\includegraphics[width=8.7cm,height=8cm]{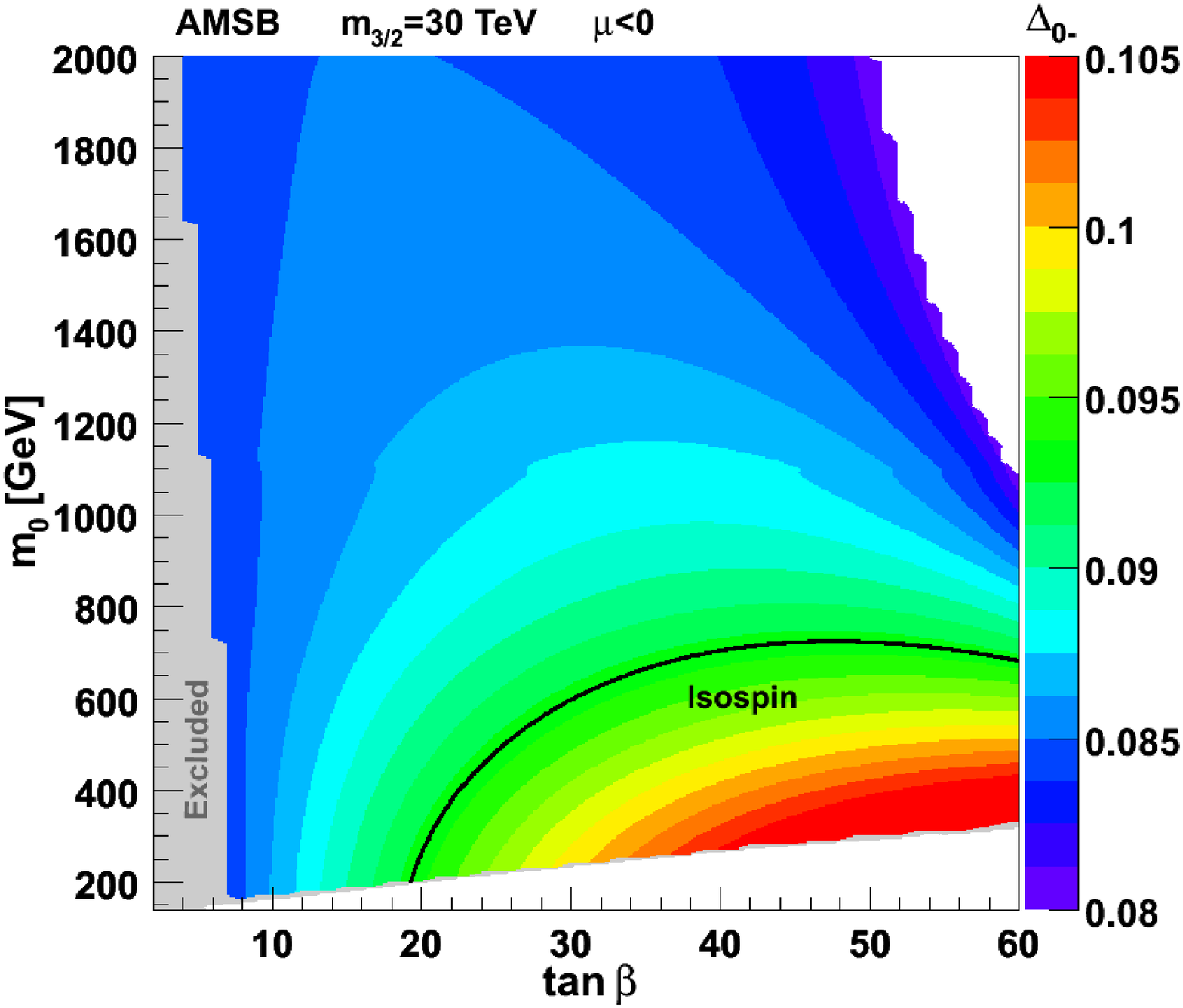}\\
\caption{Constraints on the AMSB parameter planes $(\tan\beta,m_{3/2})$ to the left, and $(\tan\beta,m_0)$ to the right. The conventions for the different regions are the same as in the precedent figures. \vspace*{0.5cm}}
\label{ambs}
\end{figure}%
We can now focus on other supersymmetry breaking scenarios, and study the influence of the isospin asymmetry for these models. First we consider the Anomaly Mediated Supersymmetry Breaking (AMSB) scenario \cite{amsb}. These mechanisms are well motivated since they preserve virtues of the gravity mediated models while the FCNC problem is solved.\\
For this scenario, we generate the masses and couplings with SOFTSUSY 2.0.14 \cite{softsusy}, and perform scans in the parameter space $\lbrace m_0$, $m_{3/2}$, $\tan\beta$, $\mathrm{sign}(\mu) \rbrace$.\\
\\
The results are presented in Fig.~\ref{ambs}, where the $(\tan\beta,m_{3/2})$ and $(\tan\beta,m_0)$ planes are studied.\\
For the $(\tan\beta,m_{3/2})$ plane, the constraints from the branching ratio are in the region already excluded by the collider mass limits. In the $(\tan\beta,m_0)$ plane, we obtain no limit from the branching ratio. However, for both cases, we obtain remarkable contours from the isospin asymmetry. This is another example in favor of investigating the isospin symmetry breaking observable.\\
\subsection{GMSB}
\begin{figure}[!t]
\begin{center}
\hspace*{-0.5cm}
\includegraphics[width=8.7cm,height=8cm]{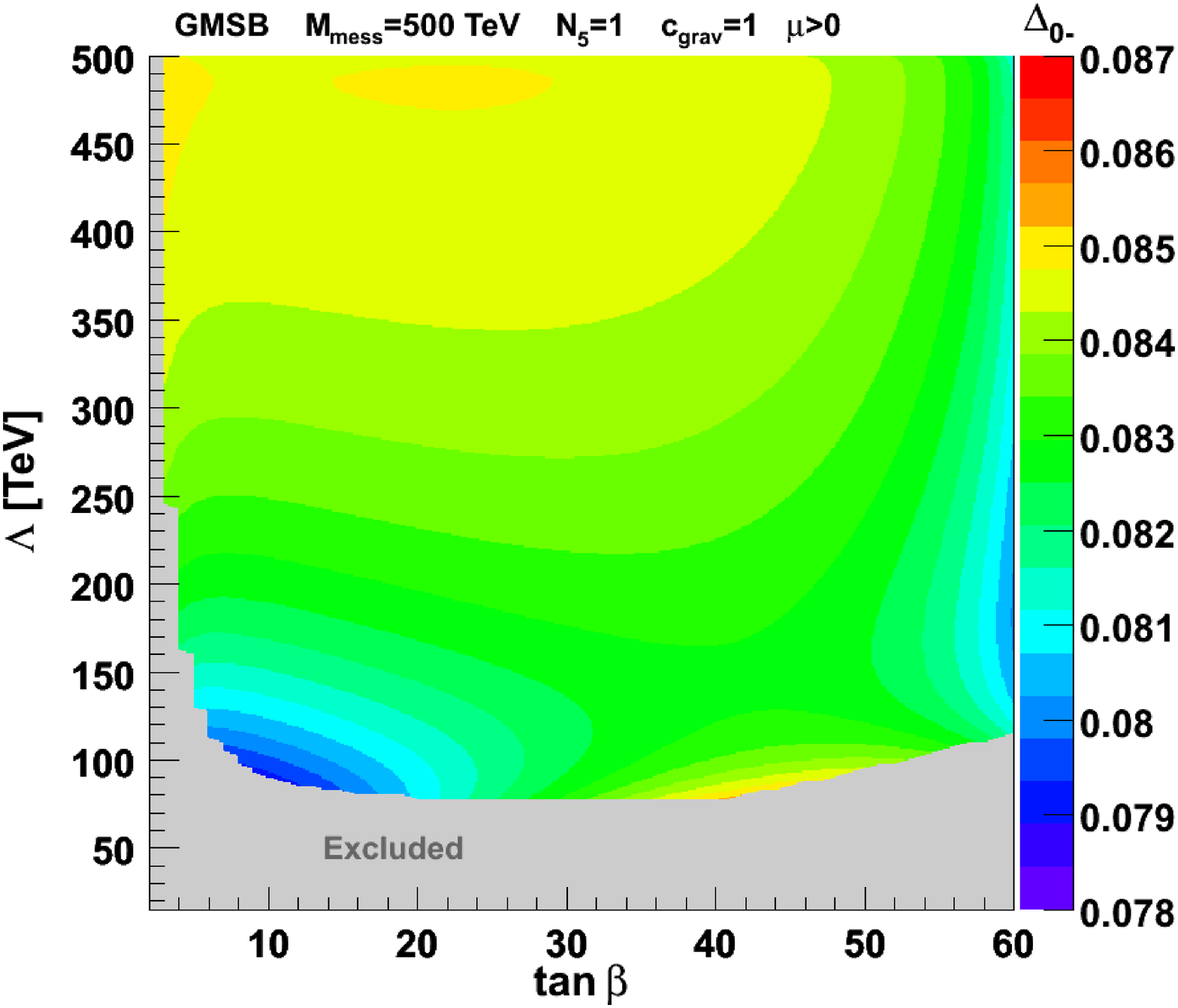}~~\includegraphics[width=8.7cm,height=8cm]{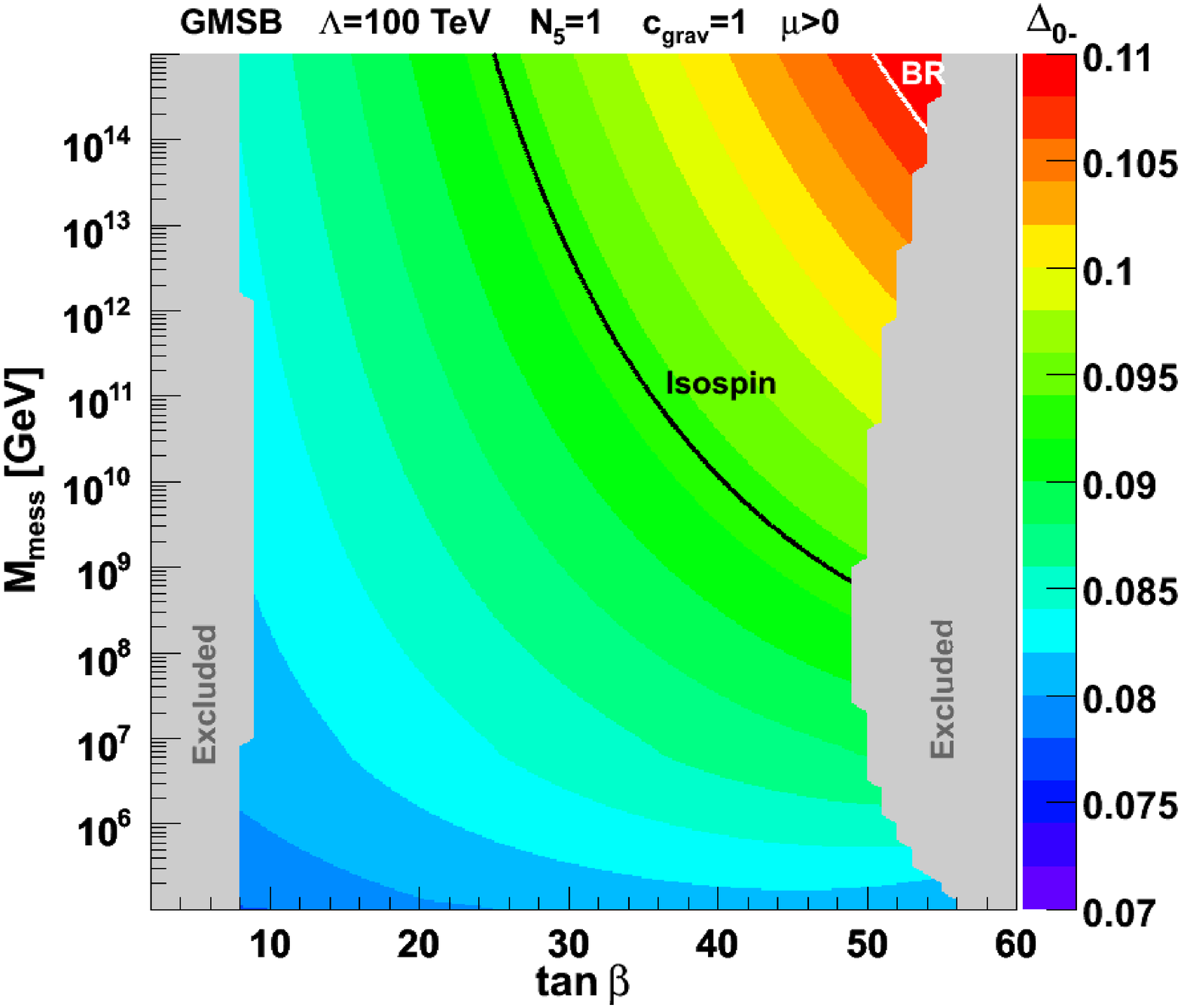}\\
\hspace*{-0.3cm}(a) \hspace*{8.2cm}(b)\\
\caption{Evolution of the isospin asymmetry in the GMSB parameter space, (a) in the plane $(\tan\beta,\Lambda)$ for $M_{\mathrm{mess}}=500$ TeV, and (b) in the plane $(\tan\beta,M_{\mathrm{mess}})$ for $\Lambda=100$ TeV. Note that in (b), the $M_{\mathrm{mess}}$ scale is logarithmic. \vspace*{0.5cm}}
\label{gmsb}
\end{center}
\end{figure}%
As a final example, we consider the Gauge Mediated Supersymmetry Breaking (GMSB) scenario \cite{gmsb}. Several regions in the parameter space $\lbrace \Lambda$, $M_{\mathrm{mess}}$, $N_5$, $c_{\mathrm{grav}}$, $\tan\beta$, $\mathrm{sign}(\mu) \rbrace$ have been investigated. Unfortunately, the available experimental data do not allow us to obtain any constraints from neither the branching ratio nor the isospin symmetry breaking for low values of the messenger scale. Indeed, in this case the stop mass is relatively large resulting in low contributions from the chargino and charged Higgs loops. Nevertheless, to show how the isospin asymmetry evolves in the GMSB parameter space for low messenger scale, we perform a scan for $M_{\mathrm{mess}} = 500$ TeV, $N_5 = 1$ and we set $c_{\mathrm{grav}} = 1$. The masses and couplings were generated with SOFTSUSY 2.0.14 \cite{softsusy}. Fig.~\ref{gmsb}a shows the results for the $(\tan\beta,\Lambda)$ plane.\\
\\
For high values of the messenger scale the situation is much better since the mixing $\tilde{t}_L - \tilde{t}_R$ is larger. Fig.~\ref{gmsb}b shows the dependence of the isospin asymmetry in function of $\tan\beta$ and  $M_{\mathrm{mess}}$. Here the isospin asymmetry starts excluding the parameters for $M_{\mathrm{mess}}$ higher than $10^{9}$ GeV and for large values of $\tan\beta$.\\
\\
With more accurate experimental data becoming available, one can hope that the isospin asymmetry could be a valuable observable even for low messenger scales.\\
\\
\subsection{Constraints on the physical masses}
Up to this point, we investigated the dependence of the isospin symmetry breaking on the parameters of different supersymmetry breaking models. We now consider the physical masses of the superpartners and investigate the values excluded by the isospin and branching ratio constraints. \\
\\
\begin{figure}[!t]
\hspace*{-0.5cm}
\includegraphics[width=9.2cm,height=8cm]{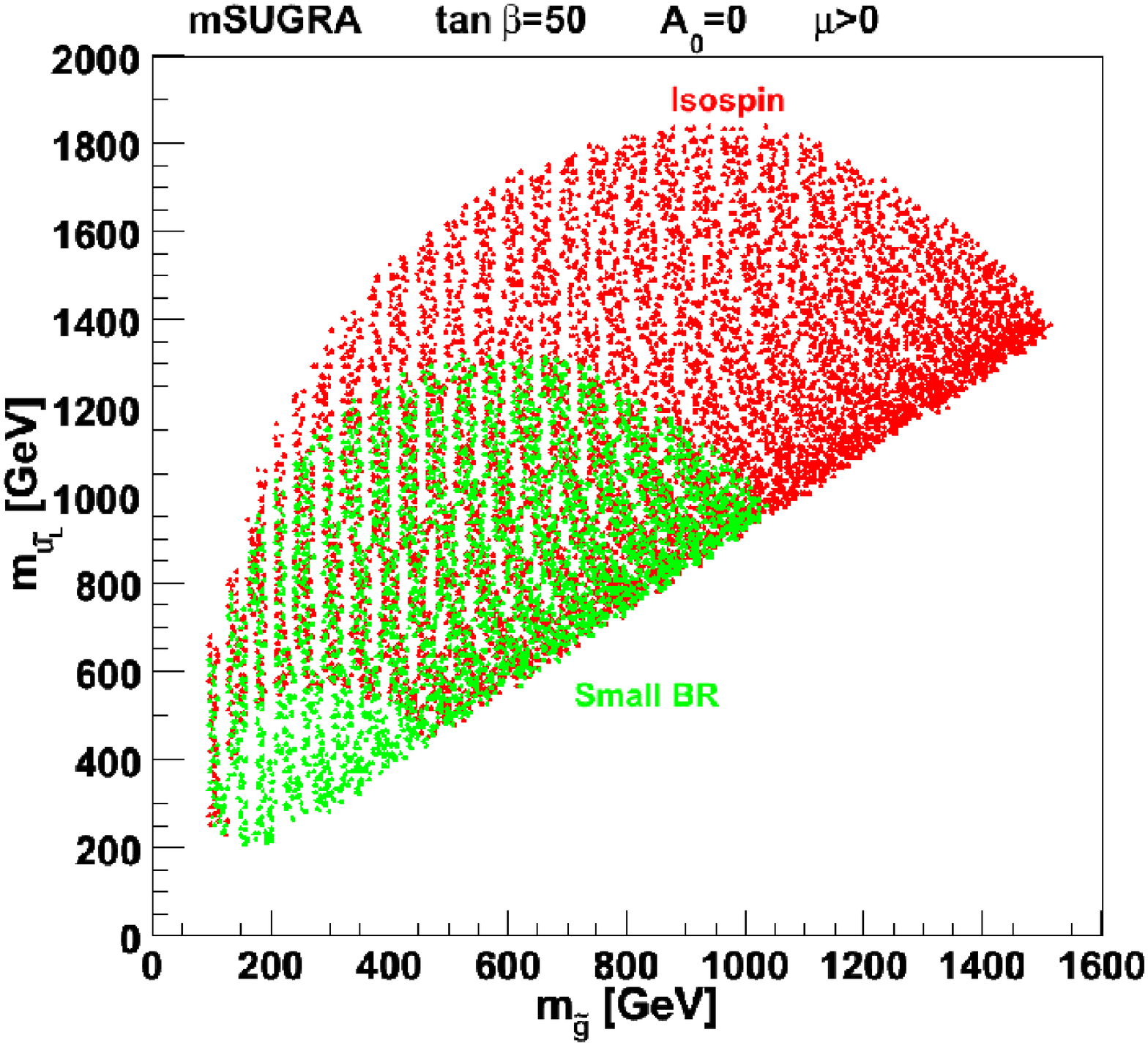}\includegraphics[width=9.2cm,height=8cm]{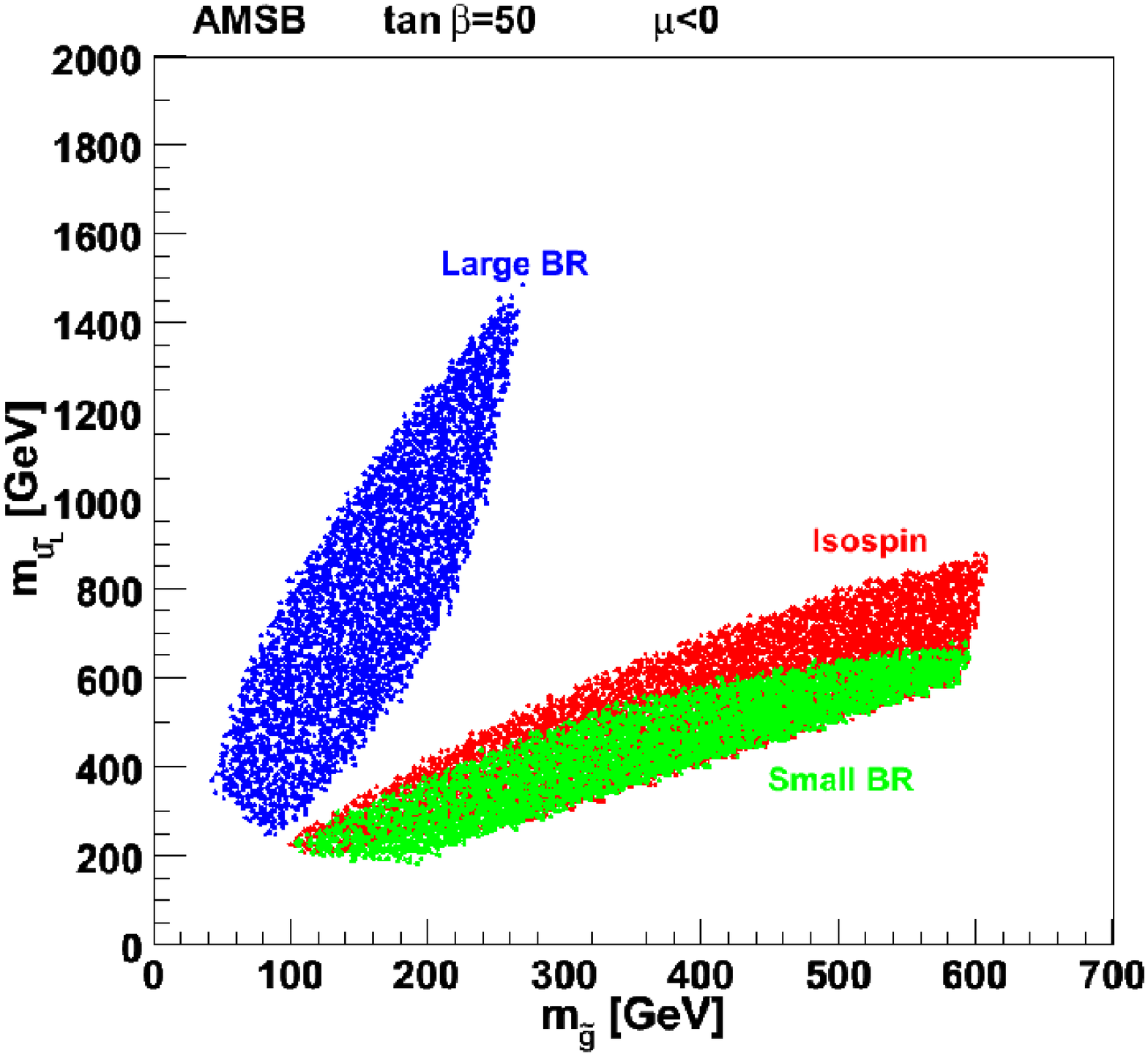}\\
\caption{The mSUGRA parameter space to the left for $\tan\beta = 50$ and $A_0=0$. To the right, the AMSB parameter space for $\tan\beta = 50$. The red dotted regions are excluded by isospin asymmetry constraints, while the green regions are excluded by the lower bound on the branching ratio and the blue region by the upper bound on the branching ratio. \vspace*{1.0cm}}
\label{qg}
\end{figure}%
We consider first the masses of gluinos and squarks, which are relevant for the strong interaction phenomenology. For this purpose, we explore mSUGRA and AMSB parameter spaces and we consider $\tilde{u}_L$ squark as an example, since all heavy squarks have approximately the same mass.
The results of the scans for ($m_{\tilde{g}}$, $m_{\tilde{u}_L}$) planes are shown in Fig.~\ref{qg}. We generated the masses and couplings using SOFTSUSY 2.0.14 \cite{softsusy}. 
The regions excluded by isospin symmetry breaking are marked with red dots, while those excluded by the branching ratio are marked with blue and green dots. The region with approximately $m_{\tilde{u}_L} > 0.8 m_{\tilde{g}}$ is not accessible due to the fact that squarks can become tachyonic at high scales in this region \cite{ellwanger}.\\
One can notice that in the AMSB parameter space, both the upper and lower bounds of the branching ratio provide restrictive constraints, but that the isospin asymmetry still rules out some additional parts of the space.\\
\\
\begin{figure}[!t]
\hspace*{-0.5cm}\includegraphics[width=8.7cm,height=8cm]{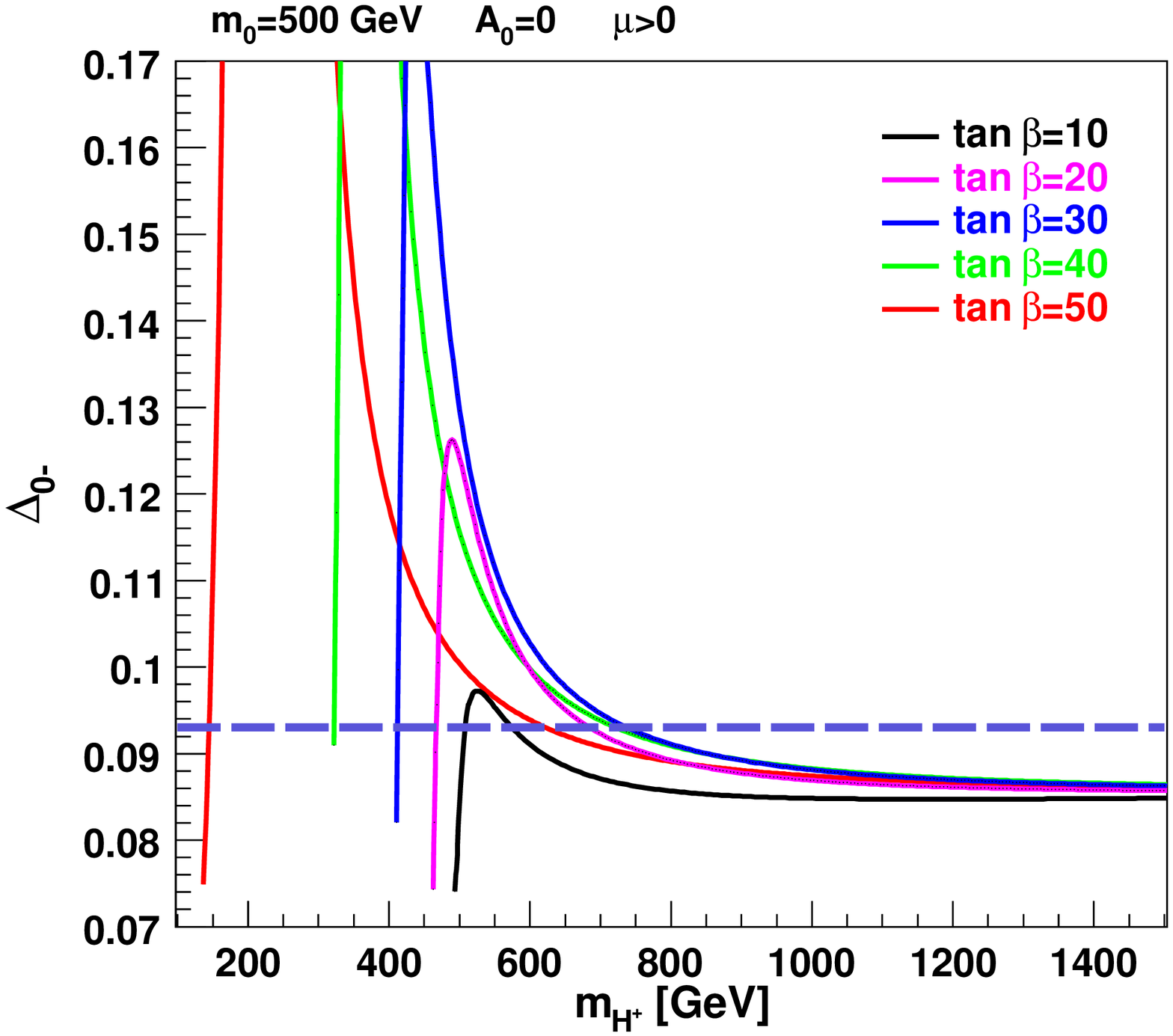}~~\includegraphics[width=8.7cm,height=8cm]{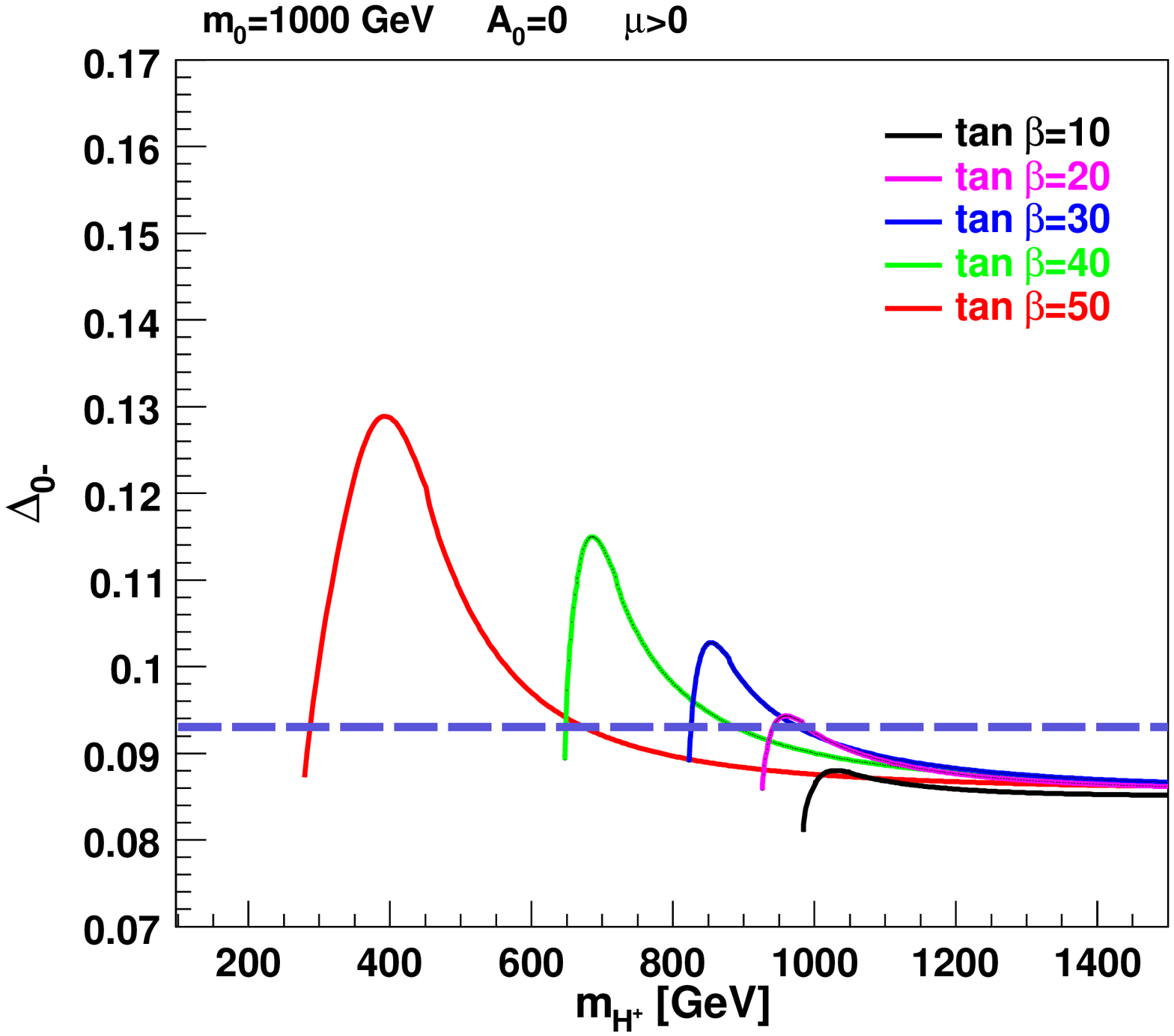}\\
\\
\caption{Constraints on the charged Higgs mass in the mSUGRA parameter space for $A_0=0$, $m_0 = 500$ GeV (left) and $m_0 = 1000$ GeV (right), and for different values of $\tan\beta$. The horizontal dashed line shows the limit from isospin asymmetry, ruling out the whole region above it. \vspace*{1.0cm}}
\label{chargedH}
\end{figure}%
As another example, we study the constraints on the charged Higgs mass. The charged Higgs boson is of special interest as a new physics discovery channel at the LHC. Fig.~\ref{chargedH} shows two-dimensional plots illustrating the constraints on the charged Higgs mass from the isospin asymmetry in the mSUGRA parameter space. The calculation has been done for different values of $\tan\beta$. The horizontal line in these plots delimits the upper bound of the allowed isospin asymmetry. One can notice that the highest restrictions are obtained for large $\tan\beta$ values. For instance, for $m_0 = 500$ GeV and $\tan\beta = 50$, the mass range between approximately 150 GeV and 630 GeV is excluded, while for $\tan\beta = 30$, the region between roughly 400 GeV and 720 GeV is excluded. One can also notice that for a higher value of $m_0$ such as $m_0 = 1000$ GeV, only larger values of charged Higgs masses are excluded.\\
\\
\section{Summary}
In this article, we explored different supersymmetric scenarios and presented the new information and constraints obtained from isospin symmetry breaking in radiative $B$ meson decays. The calculations have been performed using our recently developed program SuperIso \cite{superiso}.\\
\\
In many regions, the constraints from isospin symmetry breaking are very restrictive. Therefore, this new observable is very valuable to probe new physics scenarios. The study presented here is not exhaustive, and other regions/parameters, or models, could easily be explored by the same method. In this paper we have shown some examples of the information we can obtain using this novel observable.\\
Finally, extending this study to other models, in particular beyond the MSSM could also be of interest. \\
\\
\subsection*{Acknowledgments}
I would like to thank Sven Heinemeyer for useful discussions. I am also grateful to Gunnar Ingelman, Johan Rathsman and Oscar St\aa{}l for helpful discussions and for their comments on the manuscript.\\

\end{document}